\documentclass[aps,preprint]{revtex4}%
\usepackage{amsfonts}
\usepackage{amsmath}
\usepackage{amssymb}
\usepackage{graphicx}%
\begin{document}
\preprint{ }

\vspace*{1cm}

\begin{center}

{\Large Dilute neutron matter on the lattice}

{\Large at next-to-leading order} {\Large in chiral effective field
theory}\vspace*{0.75cm}

{Bu\={g}ra~Borasoy$^{a}$, Evgeny Epelbaum$^{b,a}$, Hermann~Krebs$^{a,b}$,
Dean~Lee$^{c,a}$, Ulf-G.~Mei{\ss }ner$^{a,b}$}\vspace*{0.75cm}

$^{a}$\textit{Helmholtz-Institut f\"{u}r Strahlen- und Kernphysik (Theorie)
Universit\"{a}t Bonn, }\linebreak\textit{Nu\ss allee 14-16, D-53115 Bonn,
Germany }

$^{b}$\textit{Institut f\"{u}r Kernphysik (Theorie), Forschungszentrum
J\"{u}lich, D-52425 J\"{u}lich, Germany }

$^{c}$\textit{Department of Physics, North Carolina State University, Raleigh,
NC 27695, USA}

\vspace*{0.75cm}

{\large Abstract}
\end{center}

We discuss lattice simulations of the ground state of dilute neutron matter at
next-to-leading order in chiral effective field theory. \ In a previous paper
the coefficients of the next-to-leading-order lattice action were determined
by matching nucleon-nucleon scattering data for momenta up to the pion mass.
\ Here the same lattice action is used to simulate the ground state of up to
12 neutrons in a periodic cube using Monte Carlo. \ We explore the density
range from 2\% to 8\% of normal nuclear density and analyze the ground state
energy as an expansion about the unitarity limit with corrections due to
finite scattering length, effective range, and $P$-wave interactions.

\pagebreak

\section{Introduction}

This is the second of a pair of papers studying chiral effective field theory
on the lattice at next-to-leading order. \ In the first paper
\cite{FirstPaper} we used nucleon-nucleon scattering data at low energies to
determine unknown operator coefficients of the next-to-leading-order lattice
action. \ We also tested model independence of the effective theory at fixed
lattice spacing by computing next-to-leading-order corrections for two
different leading-order lattice actions. \ In this paper we use the
Gaussian-smeared lattice actions LO$_{2}$ and NLO$_{2}$ defined in
\cite{FirstPaper} to simulate dilute neutron matter in a periodic cube. \ We
probe the density range from 2\% to 8\% of normal nuclear matter density.
\ Neutron-rich matter at this density is likely present in the inner crust of
neutron stars \cite{Pethick:1995di,Lattimer:2004pg}. \ The Pauli suppression
of three-body forces in dilute neutron matter makes it a good testing ground
for chiral effective field theory applied to many-nucleon systems.

The organization of the paper is as follows. \ We review the lattice
interactions contained in the leading-order (LO) and next-to-leading-order
(NLO) transfer matrices. \ These transfer matrices are rewritten in terms of
one-body interactions with auxiliary fields. \ This allows us to simulate the
ground state of the many-neutron system using transfer matrix projection and
hybrid Monte Carlo. \ The results of the simulations are compared with
published results for the ground state energy. \ We also analyze the ground
state energy as an expansion near the unitarity limit, where the scattering
length is infinite and the interactions have negligible range.

\section{Lattice transfer matrices without auxiliary fields}

In \cite{FirstPaper} we defined the lattice transfer matrix $M_{\text{LO}_{2}%
}$ at leading order and $M_{\text{NLO}_{2}}$ at next-to-leading order. \ We
use the same lattice conventions here and briefly summarize\ the relevant
definitions in the appendix. Throughout we use spatial lattice spacing
$a=(100$~MeV$)^{-1}$ and temporal lattice spacing $a_{t}=(70$~MeV$)^{-1}$.
\ We take for our physical constants $m=938.92$~MeV as the nucleon mass,
$m_{\pi}=138.08$~MeV as the pion mass, $f_{\pi}=93$~MeV as the pion decay
constant, and $g_{A}=1.26$ as the nucleon axial charge. \ In \cite{FirstPaper}
we also defined lattice actions $M_{\text{LO}_{1}}$ and $M_{\text{NLO}_{1}}$.
\ Given the significant computational resources required for the Monte Carlo
simulations, we use only the Gaussian-smeared actions $M_{\text{LO}_{2}}$ and
$M_{\text{NLO}_{2}}$. \ These yield a slightly better description of the
$S$-wave interactions expected to be dominant in dilute neutron matter.
\ Since we consider only one version of the lattice action here we drop
the\ subscript \textquotedblleft2\textquotedblright\ and write $M_{\text{LO}}$
and $M_{\text{NLO}}$. \ Tests of model independence using different lattice
actions and lattice spacings will be pursued in future studies.

The leading-order lattice transfer matrix is%
\begin{align}
M_{\text{LO}}  &  =\colon\exp\left\{  -H_{\text{free}}\alpha_{t}-\frac
{\alpha_{t}}{2L^{3}}\sum_{\vec{q}}f(q^{2})\left[  C\rho^{a^{\dag},a}(\vec
{q})\rho^{a^{\dag},a}(-\vec{q})+C_{I^{2}}\sum_{I}\rho_{I}^{a^{\dag},a}(\vec
{q})\rho_{I}^{a^{\dag},a}(-\vec{q})\right]  \right. \nonumber\\
&  +\left.  \frac{g_{A}^{2}\alpha_{t}^{2}}{8f_{\pi}^{2}q_{\pi}}\sum
_{\substack{S_{1},S_{2},I}}\sum_{\vec{n}_{1},\vec{n}_{2}}G_{S_{1}S_{2}}%
(\vec{n}_{1}-\vec{n}_{2})\rho_{S_{1},I}^{a^{\dag},a}(\vec{n}_{1})\rho
_{S_{2},I}^{a^{\dag},a}(\vec{n}_{2})\right\}  \colon. \label{LO}%
\end{align}
All of the terms appearing in Eq.~(\ref{LO}) were defined in \cite{FirstPaper}
and summarized in the appendix. \ The isospin of any two-neutron state is
$I_{z}=-1$, $I=1$. \ Therefore only the linear combination%
\begin{equation}
C^{I=1}=C+C_{I^{2}}%
\end{equation}
contributes to systems with only neutrons. \ As in \cite{FirstPaper} the
coefficient $C^{I=1}$ is set to $-3.414\times10^{-5}$ MeV$^{-2}$.

At next-to-leading order the lattice transfer matrix is%
\begin{align}
M_{\text{NLO}} &  =\colon\exp\left\{  -H_{\text{free}}\alpha_{t}-\frac
{\alpha_{t}}{2L^{3}}\sum_{\vec{q}}f(q^{2})\left[  C\rho^{a^{\dag},a}(\vec
{q})\rho^{a^{\dag},a}(-\vec{q})+C_{I^{2}}\sum_{I}\rho_{I}^{a^{\dag},a}(\vec
{q})\rho_{I}^{a^{\dag},a}(-\vec{q})\right]  \right.  \nonumber\\
&  -\left.  \alpha_{t}\left[  \Delta V+\Delta V_{I^{2}}+V_{q^{2}}%
+V_{I^{2},q^{2}}+V_{S^{2},q^{2}}+V_{S^{2},I^{2},q^{2}}+V_{(q\cdot S)^{2}%
}+V_{I^{2},(q\cdot S)^{2}}+V_{(iq\times S)\cdot k}^{I=1}\right]  \right.
\nonumber\\
&  +\left.  \frac{g_{A}^{2}\alpha_{t}^{2}}{8f_{\pi}^{2}q_{\pi}}\sum
_{\substack{S_{1},S_{2},I}}\sum_{\vec{n}_{1},\vec{n}_{2}}G_{S_{1}S_{2}}%
(\vec{n}_{1}-\vec{n}_{2})\rho_{S_{1},I}^{a^{\dag},a}(\vec{n}_{1})\rho
_{S_{2},I}^{a^{\dag},a}(\vec{n}_{2})\right\}  \colon.
\end{align}
The NLO corrections to the leading-order contact interactions are%
\begin{equation}
\Delta V=\frac{1}{2}\Delta C:\sum\limits_{\vec{n}}\rho^{a^{\dagger},a}(\vec
{n})\rho^{a^{\dagger},a}(\vec{n}):,
\end{equation}%
\begin{equation}
\Delta V_{I^{2}}=\frac{1}{2}\Delta C_{I^{2}}:\sum\limits_{\vec{n},I}\rho
_{I}^{a^{\dagger},a}(\vec{n})\rho_{I}^{a^{\dagger},a}(\vec{n}):.
\end{equation}
Again only the $I=1$ combination contributes to neutron-neutron scattering,%
\begin{equation}
\Delta C^{I=1}=\Delta C+\Delta C_{I^{2}}.
\end{equation}
There are seven independent NLO contact interactions with two derivatives,%
\begin{equation}
V_{q^{2}}=-\frac{1}{2}C_{q^{2}}:\sum\limits_{\vec{n},l}\rho^{a^{\dagger}%
,a}(\vec{n})\triangledown_{l}^{2}\rho^{a^{\dagger},a}(\vec{n}):,
\end{equation}%
\begin{equation}
V_{I^{2},q^{2}}=-\frac{1}{2}C_{I^{2},q^{2}}:\sum\limits_{\vec{n},I,l}\rho
_{I}^{a^{\dagger},a}(\vec{n})\triangledown_{l}^{2}\rho_{I}^{a^{\dagger}%
,a}(\vec{n}):,
\end{equation}%
\begin{equation}
V_{S^{2},q^{2}}=-\frac{1}{2}C_{S^{2},q^{2}}:\sum\limits_{\vec{n},S,l}\rho
_{S}^{a^{\dagger},a}(\vec{n})\triangledown_{l}^{2}\rho_{S}^{a^{\dagger}%
,a}(\vec{n}):,
\end{equation}%
\begin{equation}
V_{S^{2},I^{2},q^{2}}=-\frac{1}{2}C_{S^{2},I^{2},q^{2}}:\sum\limits_{\vec
{n},S,I,l}\rho_{S,I}^{a^{\dagger},a}(\vec{n})\triangledown_{l}^{2}\rho
_{S,I}^{a^{\dagger},a}(\vec{n}):,
\end{equation}%
\begin{equation}
V_{(q\cdot S)^{2}}=\frac{1}{2}C_{(q\cdot S)^{2}}:\sum\limits_{\vec{n}}%
\sum\limits_{S}\Delta_{S}\rho_{S}^{a^{\dagger},a}(\vec{n})\sum
\limits_{S^{\prime}}\Delta_{S^{\prime}}\rho_{S^{\prime}}^{a^{\dagger},a}%
(\vec{n}):,
\end{equation}%
\begin{equation}
V_{I^{2},(q\cdot S)^{2}}=\frac{1}{2}C_{I^{2},(q\cdot S)^{2}}:\sum
\limits_{\vec{n},I}\sum\limits_{S}\Delta_{S}\rho_{S,I}^{a^{\dagger},a}(\vec
{n})\sum\limits_{S^{\prime}}\Delta_{S^{\prime}}\rho_{S^{\prime},I}%
^{a^{\dagger},a}(\vec{n}):,
\end{equation}%
\begin{align}
V_{(iq\times S)\cdot k}^{I=1} &  =-\frac{i}{2}C_{(iq\times S)\cdot k}%
^{I=1}\left\{  \frac{3}{4}:\sum\limits_{\vec{n},l,S,l^{\prime}}\varepsilon
_{l,S,l^{\prime}}\left[  \Pi_{l}^{a^{\dagger},a}(\vec{n})\Delta_{l^{\prime}%
}\rho_{S}^{a^{\dagger},a}(\vec{n})+\Pi_{l,S}^{a^{\dagger},a}(\vec{n}%
)\Delta_{l^{\prime}}\rho^{a^{\dagger},a}(\vec{n})\right]  :\right.
\nonumber\\
&  +\left.  \frac{1}{4}:\sum\limits_{\vec{n},l,S,l^{\prime},I}\varepsilon
_{l,S,l^{\prime}}\left[  \Pi_{l,I}^{a^{\dagger},a}(\vec{n})\Delta_{l^{\prime}%
}\rho_{S,I}^{a^{\dagger},a}(\vec{n})+\Pi_{l,S,I}^{a^{\dagger},a}(\vec
{n})\Delta_{l^{\prime}}\rho_{I}^{a^{\dagger},a}(\vec{n})\right]  :\right\}  .
\end{align}
The various static densities, current densities, and symbols $\Delta_{l}$ and
$\triangledown_{l}^{2}$, are defined in the appendix. \ The $V_{(iq\times
S)\cdot k}^{I=1}$ interaction is already projected onto $I=1$. \ The other
interactions give three independent $I=1$ coefficients,%
\begin{equation}
C_{q^{2}}^{I=1}=C_{q^{2}}+C_{I^{2},q^{2}},
\end{equation}%
\begin{equation}
C_{S^{2},q^{2}}^{I=1}=C_{S^{2},q^{2}}+C_{S^{2},I^{2},q^{2}},
\end{equation}%
\begin{equation}
C_{(q\cdot S)^{2}}^{I=1}=C_{(q\cdot S)^{2}}+C_{I^{2},(q\cdot S)^{2}}.
\end{equation}

\begin{figure}
[ptb]
\begin{center}
\includegraphics[
height=4.862in,
width=4.4425in
]%
{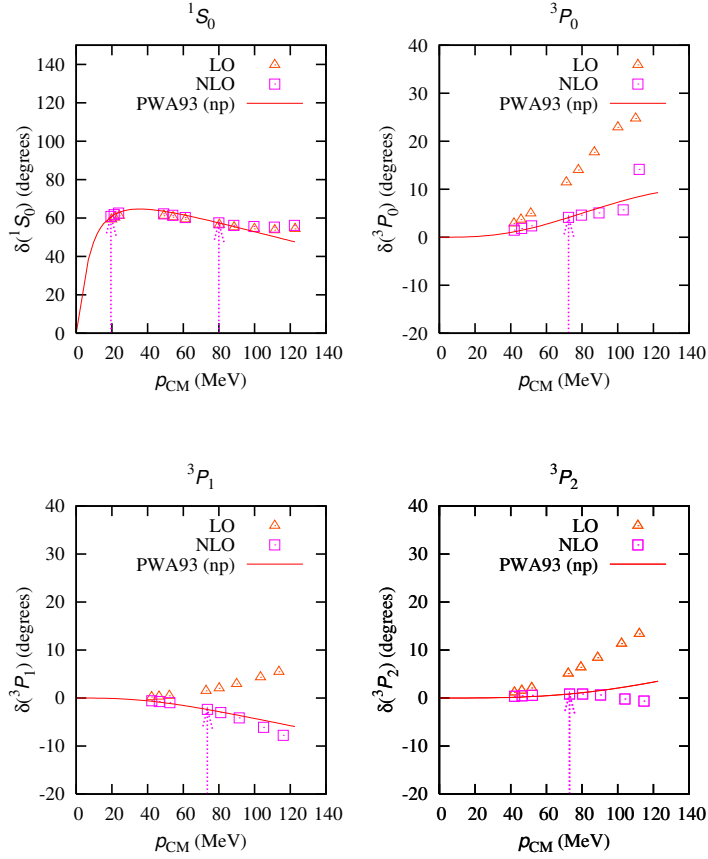}%
\caption{Scattering phase shifts for the $I=1$ singlet $S$-wave and triplet
$P$-waves versus center-of-mass momentum. \ The five arrows show data points
used to determine the five $I=1$ NLO coefficients.}%
\label{i1waves}%
\end{center}
\end{figure}

There are a total of five independent $I=1$ coefficients at NLO. \ These were
computed in \cite{FirstPaper} using the spherical wall method
\cite{Borasoy:2007vy}. \ The values of the $I=1$ coefficients are shown in
Table~\ref{NLOcoefficients}. \ \begin{table}[tb]
\caption{Results for $I=1$ NLO operator coefficients}%
\label{NLOcoefficients}%
$%
\begin{tabular}
[c]{||c|c||}\hline\hline
$\Delta C^{I=1}$ & $-7.7\times10^{-7}$ MeV$^{-2}$\\\hline
$C_{q^{2}}^{I=1}$ & $-1.42\times10^{-9}$ MeV$^{-4}$\\\hline
$C_{S^{2},q^{2}}^{I=1}$ & $-4.53\times10^{-10}$ MeV$^{-4}$\\\hline
$C_{(q\cdot S)^{2}}^{I=1}$ & $-1.80\times10^{-10}$ MeV$^{-4}$\\\hline
$C_{(iq\times S)\cdot k}^{I=1}$ & $9.81\times10^{-11}$ MeV$^{-4}%
$\\\hline\hline
\end{tabular}
$\end{table}While the LO terms in the transfer matrix are iterated
nonperturbatively, the contribution from each NLO interaction is computed
using first-order perturbation theory. \ In Fig.~\ref{i1waves} the resulting
scattering phase shifts for the $I=1$ singlet $S$-wave and triplet $P$-waves
are shown together with partial wave results from \cite{Stoks:1993tb}. \ The
five arrows show data points used to determine the five $I=1$ NLO
coefficients. \ There are also four other data points used in
\cite{FirstPaper} to determine the four $I=0$ coefficients, but these are
irrelevant for neutron-neutron scattering.%

Up until this point our lattice formalism has been constructed for a general
system of low-energy nucleons. \ For reasons of numerical efficiency for the
Monte Carlo simulation we now specialize to the case where all nucleons are
neutrons. \ With this restriction the following replacements are possible:%
\begin{equation}
C\rho^{a^{\dag},a}(\vec{q})\rho^{a^{\dag},a}(-\vec{q})+C_{I^{2}}\sum_{I}%
\rho_{I}^{a^{\dag},a}(\vec{q})\rho_{I}^{a^{\dag},a}(-\vec{q})\rightarrow
C^{I=1}\rho^{a^{\dag},a}(\vec{q})\rho^{a^{\dag},a}(-\vec{q}),
\end{equation}%
\begin{equation}
\sum_{\substack{S_{1},S_{2},I}}\sum_{\vec{n}_{1},\vec{n}_{2}}G_{S_{1}S_{2}%
}(\vec{n}_{1}-\vec{n}_{2})\rho_{S_{1},I}^{a^{\dag},a}(\vec{n}_{1})\rho
_{S_{2},I}^{a^{\dag},a}(\vec{n}_{2})\rightarrow\sum_{S_{1},S_{2}}\sum_{\vec
{n}_{1},\vec{n}_{2}}G_{S_{1}S_{2}}(\vec{n}_{1}-\vec{n}_{2})\rho_{S_{1}%
}^{a^{\dag},a}(\vec{n}_{1})\rho_{S_{2}}^{a^{\dag},a}(\vec{n}_{2}).
\end{equation}
This change has no effect on the interactions between neutrons but leads to
the simplified transfer matrix,%
\begin{align}
M_{\text{LO}}  &  \rightarrow\colon\exp\left\{  -H_{\text{free}}\alpha
_{t}-\frac{C^{I=1}\alpha_{t}}{2L^{3}}\sum_{\vec{q}}f(q^{2})\rho^{a^{\dag}%
,a}(\vec{q})\rho^{a^{\dag},a}(-\vec{q})\right. \nonumber\\
&  +\left.  \frac{g_{A}^{2}\alpha_{t}^{2}}{8f_{\pi}^{2}q_{\pi}}\sum
_{\substack{S_{1},S_{2}}}\sum_{\vec{n}_{1},\vec{n}_{2}}G_{S_{1}S_{2}}(\vec
{n}_{1}-\vec{n}_{2})\rho_{S_{1}}^{a^{\dag},a}(\vec{n}_{1})\rho_{S_{2}%
}^{a^{\dag},a}(\vec{n}_{2})\right\}  \colon. \label{MLO}%
\end{align}
At next-to-leading order the simplified neutron transfer matrix is%
\begin{align}
M_{\text{NLO}}  &  \rightarrow\colon\exp\left\{  -H_{\text{free}}\alpha
_{t}-\frac{C^{I=1}\alpha_{t}}{2L^{3}}\sum_{\vec{q}}f(q^{2})\rho^{a^{\dag}%
,a}(\vec{q})\rho^{a^{\dag},a}(-\vec{q})\right. \nonumber\\
&  -\left.  \alpha_{t}\left[  \frac{\Delta C^{I=1}}{\Delta C}\Delta
V+\frac{C_{q^{2}}^{I=1}}{C_{q^{2}}}V_{q^{2}}+\frac{C_{S^{2},q^{2}}^{I=1}%
}{C_{S^{2},q^{2}}}V_{S^{2},q^{2}}+\frac{C_{(q\cdot S)^{2}}^{I=1}}{C_{(q\cdot
S)^{2}}}V_{(q\cdot S)^{2}}+V_{(iq\times S)\cdot k}^{I=1}\right]  \right.
\nonumber\\
&  +\left.  \frac{g_{A}^{2}\alpha_{t}^{2}}{8f_{\pi}^{2}q_{\pi}}\sum
_{S_{1},S_{2},}\sum_{\vec{n}_{1},\vec{n}_{2}}G_{S_{1}S_{2}}(\vec{n}_{1}%
-\vec{n}_{2})\rho_{S_{1}}^{a^{\dag},a}(\vec{n}_{1})\rho_{S_{2}}^{a^{\dag}%
,a}(\vec{n}_{2})\right\}  \colon. \label{MNLO}%
\end{align}
In the following we use these simplified forms for the leading-order and
next-to-leading-order transfer matrices.

\section{Lattice transfer matrices with auxiliary fields}

The transfer matrices in Eq.~(\ref{MLO}) and (\ref{MNLO}) can be rewritten as
one-body interactions with auxiliary fields. \ This auxiliary-field
formulation is useful for the many-body simulation. \ The exact equivalence
between lattice formalisms with and without auxiliary fields was shown in
\cite{Lee:2006hr,Borasoy:2006qn}. \ We summarize the results here.

In neutron-neutron scattering only the neutral pion contributes to one-pion
exchange. \ Up to this point we have been writing the rescaled neutral pion
field as $\pi_{3}^{\prime}$. \ In the following we drop the subscript
\textquotedblleft3\textquotedblright\ and simply write $\pi^{\prime}$. \ Let
$M^{(n_{t})}(\pi^{\prime},s)$ be the leading-order\ auxiliary-field transfer
matrix at time step $n_{t}$,%
\begin{align}
M^{(n_{t})}(\pi^{\prime},s)  &  =\colon\exp\left\{  -H_{\text{free}}\alpha
_{t}+\frac{g_{A}\alpha_{t}}{2f_{\pi}\sqrt{q_{\pi}}}%
{\displaystyle\sum_{S}}
\Delta_{S}\pi^{\prime}(\vec{n},n_{t})\rho_{S}^{a^{\dag},a}(\vec{n})\right.
\nonumber\\
&  \qquad\qquad\left.  +\sqrt{-C^{I=1}\alpha_{t}}\sum_{\vec{n}}s(\vec{n}%
,n_{t})\rho^{a^{\dag},a}(\vec{n})\right\}  \colon.
\end{align}
We can write $M_{\text{LO}}$ as the normalized integral%
\begin{equation}
M_{\text{LO}}=\frac{%
{\displaystyle\int}
D\pi^{\prime}Ds\;e^{-S_{\pi\pi}^{(n_{t})}-S_{ss}^{(n_{t})}}M^{(n_{t})}%
(\pi^{\prime},s)}{%
{\displaystyle\int}
D\pi^{\prime}Ds\;e^{-S_{\pi\pi}^{(n_{t})}-S_{ss}^{(n_{t})}}}, \label{LOaux}%
\end{equation}
where $S_{\pi\pi}^{(n_{t})}$ is the piece of the instantaneous pion action\ in
Eq.~(\ref{pionaction}) containing the neutral pion field at time step $n_{t}$,%
\begin{equation}
S_{\pi\pi}^{(n_{t})}(\pi^{\prime})=\frac{1}{2}\sum_{\vec{n}}\pi^{\prime}%
(\vec{n},n_{t})\pi^{\prime}(\vec{n},n_{t})-\frac{\alpha_{t}}{q_{\pi}}%
\sum_{\vec{n},l}\pi^{\prime}(\vec{n},n_{t})\pi^{\prime}(\vec{n}+\hat{l}%
,n_{t}),
\end{equation}
and $S_{ss}^{(n_{t})}$ is the auxiliary-field action at time step $n_{t}$,%
\begin{equation}
S_{ss}^{(n_{t})}=\frac{1}{2}\sum_{\vec{n},\vec{n}^{\prime}}s(\vec{n}%
,n_{t})f^{-1}(\vec{n}-\vec{n}^{\prime})s(\vec{n}^{\prime},n_{t}),
\end{equation}
with
\begin{equation}
f^{-1}(\vec{n}-\vec{n}^{\prime})=\frac{1}{L^{3}}\sum_{\vec{q}}\frac{1}%
{f(q^{2})}e^{-i\vec{q}\cdot(\vec{n}-\vec{n}^{\prime})}.
\end{equation}

The NLO interactions require several additional auxiliary fields. \ Let%
\begin{align}
U^{(n_{t})}(\varepsilon)  &  =\sum_{\vec{n}}\varepsilon_{\rho}(\vec{n}%
,n_{t})\rho^{a^{\dagger},a}(\vec{n})+\sum_{\vec{n},S}\varepsilon_{\rho_{S}%
}(\vec{n},n_{t})\rho_{S}^{a^{\dagger},a}(\vec{n})+\sum_{\vec{n},S}%
\varepsilon_{\Delta_{S}\rho}(\vec{n},n_{t})\Delta_{S}\rho^{a^{\dagger},a}%
(\vec{n})\nonumber\\
&  +\sum_{\vec{n},S,S^{\prime}}\varepsilon_{\Delta_{S}\rho_{S^{\prime}}}%
(\vec{n},n_{t})\Delta_{S}\rho_{S^{\prime}}^{a^{\dagger},a}(\vec{n})+\sum
_{\vec{n},l}\varepsilon_{\triangledown_{l}^{2}\rho}(\vec{n},n_{t}%
)\triangledown_{l}^{2}\rho^{a^{\dagger},a}(\vec{n})\nonumber\\
&  +\sum_{\vec{n},l,S}\varepsilon_{\triangledown_{l}^{2}\rho_{S}}(\vec
{n},n_{t})\triangledown_{l}^{2}\rho_{S}^{a^{\dagger},a}(\vec{n})+\sum_{\vec
{n},l}\varepsilon_{\Pi_{l}}(\vec{n},n_{t})\Pi_{l}^{a^{\dagger},a}(\vec
{n})+\sum_{\vec{n},l,S}\varepsilon_{\Pi_{l,S}}(\vec{n},n_{t})\Pi
_{l,S}^{a^{\dagger},a}(\vec{n}).
\end{align}
With these extra fields and linear functional $U^{(n_{t})}(\varepsilon)$ we
define%
\begin{align}
M^{(n_{t})}(\pi^{\prime},s,\varepsilon)  &  =\colon\exp\left\{
-H_{\text{free}}\alpha_{t}+\frac{g_{A}\alpha_{t}}{2f_{\pi}\sqrt{q_{\pi}}}%
{\displaystyle\sum_{S}}
\Delta_{S}\pi^{\prime}(\vec{n},n_{t})\rho_{S}^{a^{\dag},a}(\vec{n})\right.
\nonumber\\
&  \qquad\qquad\left.  +\sqrt{-C^{I=1}\alpha_{t}}\sum_{\vec{n}}s(\vec{n}%
,n_{t})\rho^{a^{\dag},a}(\vec{n})+\sqrt{\alpha_{t}}U^{(n_{t})}(\varepsilon
)\right\}  \colon.
\end{align}
We also define the normalized integral,%
\begin{equation}
M^{(n_{t})}(\varepsilon)=\frac{%
{\displaystyle\int}
D\pi^{\prime}Ds\;e^{-S_{\pi\pi}^{(n_{t})}-S_{ss}^{(n_{t})}}M^{(n_{t})}%
(\pi^{\prime},s,\varepsilon)}{%
{\displaystyle\int}
D\pi^{\prime}Ds\;e^{-S_{\pi\pi}^{(n_{t})}-S_{ss}^{(n_{t})}}}.
\label{LOaux_eps}%
\end{equation}
When all $\varepsilon$ fields are set to zero we recover $M_{\text{LO}}$,%
\begin{equation}
M^{(n_{t})}(0)=M_{\text{LO}}\text{.}%
\end{equation}
To first order the NLO interactions in $M_{\text{NLO}}$ can be written as a
sum of bilinear derivatives of $M^{(n_{t})}(\varepsilon)$ with respect to the
$\varepsilon$ fields at $\varepsilon=0$,%
\begin{align}
M_{\text{NLO}}  &  =M_{\text{LO}}\nonumber\\
&  -\frac{1}{2}\Delta C^{I=1}\sum_{\vec{n}}\left.  \frac{\delta}%
{\delta\varepsilon_{\rho}(\vec{n},n_{t})}\frac{\delta}{\delta\varepsilon
_{\rho}(\vec{n},n_{t})}M^{(n_{t})}(\varepsilon)\right\vert _{\varepsilon
=0}\nonumber\\
&  +\frac{1}{2}C_{q^{2}}^{I=1}\sum_{\vec{n}}\left.  \frac{\delta}%
{\delta\varepsilon_{\rho}(\vec{n},n_{t})}\frac{\delta}{\delta\varepsilon
_{\triangledown_{l}^{2}\rho}(\vec{n},n_{t})}M^{(n_{t})}(\varepsilon
)\right\vert _{\varepsilon=0}+\;\cdots.
\end{align}

\section{Transfer matrix projection method}

We use the transfer matrix projection method introduced in \cite{Lee:2005fk}.
\ First we give a short overview using simple continuum notation. \ Let
$\left\vert \Psi^{\text{free}}\right\rangle $ be a Slater determinant of
free-particle standing waves in a periodic cube for $N$ neutrons. \ Let
$H_{\text{LO}}$ be the Hamiltonian at leading order, and $H_{\text{NLO}}$ be
the Hamiltonian at next-to-leading order. \ Let $H_{\text{SU(2)}\not \pi }$ be
the same as $H_{\text{LO}}$, but with one-pion exchange turned off by setting
$g_{A}$ to zero. \ As the notation suggests, $H_{\text{SU(2)}\not \pi }$ is
invariant under an exact SU(2) intrinsic-spin symmetry.

Let us define a trial wavefunction%
\begin{equation}
\left\vert \Psi(t^{\prime})\right\rangle =\exp\left(  -H_{\text{SU}%
(2)\not \pi }t^{\prime}\right)  \left\vert \Psi^{\text{free}}\right\rangle .
\end{equation}
In this approach $\exp\left(  -H_{\text{SU(2)}\not \pi }t^{\prime}\right)  $
acts as an approximate low-energy filter. \ In the auxiliary-field Monte Carlo
calculation this part of the Euclidean time propagation is positive definite
for any even number of neutrons invariant under the SU(2) intrinsic-spin
symmetry \cite{Lee:2004hc,Chen:2004rq,Lee:2007eu}. \ With this trial
wavefunction we define the amplitude,%
\begin{equation}
Z(t)=\left\langle \Psi(t^{\prime})\right\vert \exp\left(  -H_{\text{LO}%
}t\right)  \left\vert \Psi(t^{\prime})\right\rangle ,
\end{equation}
as well as the transient energy,%
\begin{equation}
E_{\text{LO}}(t)=-\frac{\partial}{\partial t}\left[  \ln Z(t)\right]  .
\end{equation}
In limit of large Euclidean time $t$ we get%
\begin{equation}
\lim_{t\rightarrow\infty}E_{\text{LO}}(t)=E_{0,\text{LO}},
\end{equation}
where $E_{0,\text{LO}}$ is the energy of the lowest eigenstate $\left\vert
\Psi_{0}\right\rangle $ of $H_{\text{LO}}$ with nonzero inner product with
$\left\vert \Psi(t^{\prime})\right\rangle $.

To compute the expectation value of some operator $O$ we define%
\begin{equation}
Z_{O}(t)=\left\langle \Psi(t^{\prime})\right\vert \exp\left(  -H_{\text{LO}%
}t/2\right)  O\,\exp\left(  -H_{\text{LO}}t/2\right)  \left\vert
\Psi(t^{\prime})\right\rangle .
\end{equation}
The expectation value of $O$ for $\left\vert \Psi_{0}\right\rangle $ is given
by the large $t$ limit,%
\begin{equation}
\lim_{t\rightarrow\infty}\frac{Z_{O}(t)}{Z(t)}=\left\langle \Psi
_{0}\right\vert O\left\vert \Psi_{0}\right\rangle .
\end{equation}
Let $H_{\text{NLO}}$ be the next-to-leading-order Hamiltonian. \ Corrections
to the energy at next-to-leading order can be computed using $O=H_{\text{NLO}%
}-H_{\text{LO}}$. \ Then%
\begin{equation}
\lim_{t\rightarrow\infty}\frac{Z_{O}(t)}{Z(t)}=E_{0,\text{NLO}}-E_{0,\text{LO}%
},
\end{equation}
where $E_{0,\text{NLO}}$ is the ground state energy at next-to-leading order.

On the lattice we construct $\left\vert \Psi(t^{\prime})\right\rangle $ using%
\begin{equation}
\left\vert \Psi(t^{\prime})\right\rangle =\left(  M_{\text{SU(2)}\not \pi
}\right)  ^{L_{t_{o}}}\left\vert \Psi^{\text{free}}\right\rangle ,
\end{equation}
where $t^{\prime}=L_{t_{o}}\alpha_{t}$ and $L_{t_{o}}$ is the number of
\textquotedblleft outer\textquotedblright\ time steps. \ The amplitude $Z(t)$
is constructed using%
\begin{equation}
Z(t)=\left\langle \Psi(t^{\prime})\right\vert \left(  M_{\text{LO}}\right)
^{L_{t_{i}}}\left\vert \Psi(t^{\prime})\right\rangle ,
\end{equation}
where $t=L_{t_{i}}\alpha_{t}$ and $L_{t_{i}}$ is the number of
\textquotedblleft inner\textquotedblright\ time steps. \ The transient energy%
\begin{equation}
E_{\text{LO}}(t+\alpha_{t}/2)
\end{equation}
is given by the ratio of the amplitudes for $t$ and $t+\alpha_{t}$,%
\begin{equation}
e^{-E_{\text{LO}}(t+\alpha_{t}/2)\cdot\alpha_{t}}=\frac{Z(t+\alpha_{t})}%
{Z(t)}.
\end{equation}
The ground state energy $E_{0,\text{LO}}$ equals the asymptotic limit,%
\begin{equation}
E_{0,\text{LO}}=\lim_{t\rightarrow\infty}E_{\text{LO}}(t+\alpha_{t}/2).
\end{equation}

For the ground state energy at NLO we compute expectation values of
$M_{\text{NLO}}$ and $M_{\text{LO}}$ inserted in the middle of a string of LO
transfer matrices,%
\begin{equation}
Z_{M_{\text{NLO}}}(t)=\left\langle \Psi(t^{\prime})\right\vert \left(
M_{\text{LO}}\right)  ^{L_{t_{i}}/2}M_{\text{NLO}}\left(  M_{\text{LO}%
}\right)  ^{L_{t_{i}}/2}\left\vert \Psi(t^{\prime})\right\rangle ,
\end{equation}%
\begin{equation}
Z_{M_{\text{LO}}}(t)=\left\langle \Psi(t^{\prime})\right\vert \left(
M_{\text{LO}}\right)  ^{L_{t_{i}}/2}M_{\text{LO}}\left(  M_{\text{LO}}\right)
^{L_{t_{i}}/2}\left\vert \Psi(t^{\prime})\right\rangle .
\end{equation}
Clearly $Z_{M_{\text{LO}}}(t)$ is the same as $Z(t+\alpha_{t})$. \ We use the
ratio of amplitudes,%
\begin{equation}
\frac{Z_{M_{\text{NLO}}}(t)}{Z_{M_{\text{LO}}}(t)}=1-\Delta E_{\text{NLO}%
}(t)\alpha_{t}+\cdots,
\end{equation}
to define the transient NLO energy correction $\Delta E_{\text{NLO}}(t)$.
\ The ellipsis denotes terms which are beyond first order in the
NLO\ coefficients. \ The NLO ground state energy $E_{0,\text{NLO}}$ is
calculated using%
\begin{equation}
E_{0,\text{NLO}}=E_{0,\text{LO}}+\lim_{t\rightarrow\infty}\Delta
E_{0,\text{NLO}}(t),
\end{equation}

The Monte Carlo simulation is carried out using the auxiliary-field
formulations of the transfer matrices. \ Integrations over auxiliary and pion
field configurations are computed using hybrid Monte Carlo with endpoint
importance sampling. \ Details of this method can be found in the literature
\cite{Lee:2005fk,Lee:2006hr,Borasoy:2006qn}.

\section{Precision tests}

We use systems of two neutrons to test the auxiliary-field Monte Carlo
simulations. \ We calculate the same observables using both the Monte Carlo
code and the exact transfer matrix without auxiliary fields. \ We choose a
small system so that stochastic errors are small enough to expose disagreement
at the $0.1\%-1\%$ level. \ We choose the spatial length of lattice to be
$L=4$ and set the outer time steps $L_{t_{o}}=2$ and inner time steps
$L_{t_{i}}=4$.

For the first test we choose $\left\vert \Psi^{\text{free}}\right\rangle $ to
be a spin-singlet state built from the Slater determinant of standing waves
$\left\vert \psi_{1}\right\rangle $ and $\left\vert \psi_{2}\right\rangle $
with%
\begin{equation}
\left\langle 0\right\vert a_{i,j}(\vec{n})\left\vert \psi_{1}\right\rangle
\propto\delta_{i,0}\delta_{j,1},\qquad\left\langle 0\right\vert a_{i,j}%
(\vec{n})\left\vert \psi_{2}\right\rangle \propto\delta_{i,1}\delta_{j,1}.
\end{equation}
For the second test we choose a spin-triplet state with standing waves%
\begin{equation}
\left\langle 0\right\vert a_{i,j}(\vec{n})\left\vert \psi_{1}\right\rangle
\propto\cos(\tfrac{2\pi n_{1}}{L})\delta_{i,0}\delta_{j,1},\qquad\left\langle
0\right\vert a_{i,j}(\vec{n})\left\vert \psi_{2}\right\rangle \propto
\sin(\tfrac{2\pi n_{1}}{L})\delta_{i,0}\delta_{j,1}.
\end{equation}
Comparisons between Monte Carlo results (MC) and exact transfer matrix
calculations (exact) are shown in Table \ref{precision}. \ The numbers in
parentheses are the estimated stochastic errors. \ \begin{table}[tb]
\caption{Monte Carlo results versus exact transfer matrix calculations for the
two-neutron spin singlet $S=0$ and spin triplet $S=1.$}%
\label{precision}
\begin{tabular}
[c]{||c|c|c|c|c||}\hline\hline
& $S=0$ (MC) & $S=0$ (exact) & $S=1$ (MC) & $S=1$ (exact)\\\hline
$E_{\text{LO}}(t+\alpha_{t}/2)$ [MeV] & $-2.93(2)$ & $-2.9242$ & $24.99(10)$ &
$25.030$\\\hline
$\frac{\partial\left(  \Delta E_{\text{NLO}}(t)\right)  }{\partial\left(
\Delta C^{I=1}\right)  }$ [$10^{4}$ MeV$^{3}$] & $4.869(6)$ & $4.8620$ &
$0.0003(2)$ & $0$\\\hline
$\frac{\partial\left(  \Delta E_{\text{NLO}}(t)\right)  }{\partial\left(
C_{q^{2}}^{I=1}\right)  }$ [$10^{9}$ MeV$^{5}$] & $1.617(3)$ & $1.6140$ &
$-1.853(4)$ & $-1.8524$\\\hline
$\frac{\partial\left(  \Delta E_{\text{NLO}}(t)\right)  }{\partial\left(
C_{S^{2},q^{2}}^{I=1}\right)  }$ [$10^{9}$ MeV$^{5}$] & $-4.85(1)$ & $-4.8419$
& $-1.851(4)$ & $-1.8524$\\\hline
$\frac{\partial\left(  \Delta E_{\text{NLO}}(t)\right)  }{\partial\left(
C_{(q\cdot S)^{2}}^{I=1}\right)  }$ [$10^{8}$ MeV$^{5}$] & $-6.00(1)$ &
$-5.9822$ & $7.00(2)$ & $7.0012$\\\hline
$\frac{\partial\left(  \Delta E_{\text{NLO}}(t)\right)  }{\partial\left(
C_{(iq\times S)\cdot k}^{I=1}\right)  }$ [$10^{7}$ MeV$^{5}$] & $0.011(7)$ &
$0$ & $7.8(1)$ & $7.8743$\\\hline
$\Delta E_{\text{NLO}}(t)$ [MeV] & $-0.0252(3)$ & $-0.025025$ & $3.349(7)$ &
$3.3490$\\\hline\hline
\end{tabular}
\end{table}We see that in each case the agreement is comparable to the
estimated stochastic error.

\section{Results}

We simulate $N=8$ and $N=12$ neutrons on periodic cube lattices with spatial
length $L=5,6,7$ lattice units. \ For each value of $N$ and $L$ we set
$L_{t_{o}}=10$ and vary $L_{t_{i}}$ from $2$ to $12$. \ For $\left\vert
\Psi^{\text{free}}\right\rangle $ we take the Slater determinant formed by
standing waves%
\begin{equation}
\left\langle 0\right\vert a_{i,j}(\vec{n})\left\vert \psi_{2k+1}\right\rangle
\propto f_{k}(\vec{n})\delta_{i,0}\delta_{j,1},\qquad\left\langle 0\right\vert
a_{i,j}(\vec{n})\left\vert \psi_{2k+2}\right\rangle \propto f_{k}(\vec
{n})\delta_{i,1}\delta_{j,1},
\end{equation}
where%
\[
f_{0}(\vec{n})=1,\quad f_{1}(\vec{n})=\cos(\tfrac{2\pi n_{1}}{L}),\quad
f_{2}(\vec{n})=\sin(\tfrac{2\pi n_{1}}{L}),
\]%
\begin{equation}
f_{3}(\vec{n})=\cos(\tfrac{2\pi n_{2}}{L}),\quad f_{4}(\vec{n})=\sin
(\tfrac{2\pi n_{2}}{L}),\quad f_{5}(\vec{n})=\cos(\tfrac{2\pi n_{3}}{L}).
\end{equation}
For $N=8$ we use $k=0,1,\cdots,3,$ and for $N=12$ we take $k=0,1,\cdots,5$.
\ For each value of $L_{t_{i}}$ a total of about $10^{6}$ hybrid Monte Carlo
trajectories are generated by $1024$ processors, each running completely
independent trajectories. \ Averages and stochastic errors are computed by
comparing the results of all $1024$ processors. \ 

\begin{figure}
[ptb]
\begin{center}
\includegraphics[
height=2.6896in,
width=5.0289in
]%
{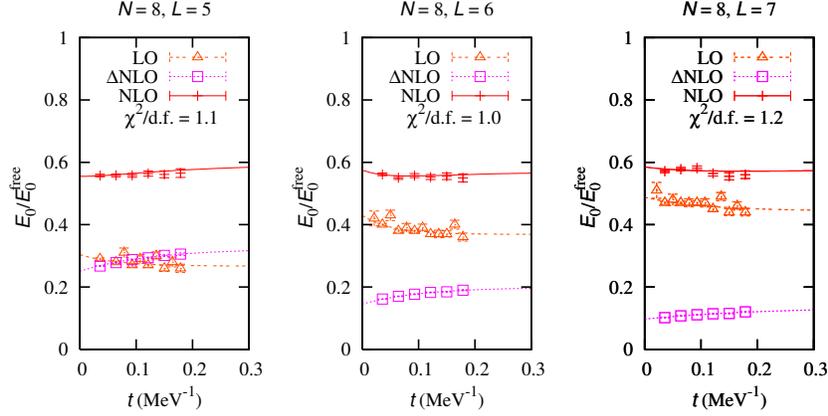}%
\caption{Plots of the three energy ratios defined in Eq.~(\ref{ratios}) for
$N=8$ and $L=5,6,7$. \ These are labelled as LO, $\Delta$NLO, NLO
respectively.}%
\label{n8}%
\end{center}
\end{figure}

\begin{figure}
[ptbptb]
\begin{center}
\includegraphics[
height=2.6887in,
width=5.0298in
]%
{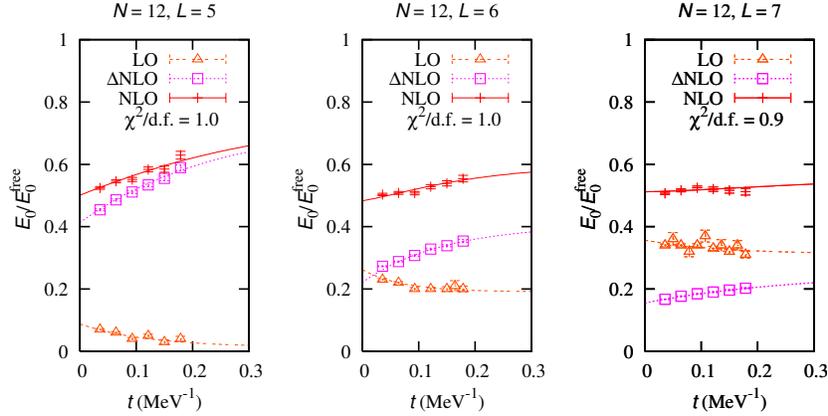}%
\caption{Plots of the three energy ratios defined in Eq.~(\ref{ratios}) for
$N=12$ and $L=5,6,7$. \ These are labelled as LO, $\Delta$NLO, NLO
respectively.}%
\label{n12}%
\end{center}
\end{figure}

Let $E_{0}^{\text{free}}$ be the energy of the ground state for noninteracting
neutrons. \ In Fig.~\ref{n8} we show the dimensionless ratios%
\begin{equation}
\frac{E_{\text{LO}}(t)}{E_{0}^{\text{free}}}\text{,\quad}\frac{\Delta
E_{\text{NLO}}(t)}{E_{0}^{\text{free}}},\text{\quad}\frac{E_{\text{LO}%
}(t)+\Delta E_{\text{NLO}}(t)}{E_{0}^{\text{free}}}\text{,} \label{ratios}%
\end{equation}
versus Euclidean time $t$ for $N=8$ and $L=5,6,7$. \ These are labelled using
the shorthand LO, $\Delta$NLO, and NLO respectively. \ The same quantities are
shown in Fig.~\ref{n12} for $N=12$. \ The lattice calculations for $\Delta
E_{\text{NLO}}(t)$ require an even number of time steps and so fewer data
points are available. \ In addition to the Monte Carlo data we plot the
asymptotic expressions,%
\begin{equation}
\frac{E_{\text{LO}}(t)}{E_{0}^{\text{free}}}\approx\frac{E_{0,\text{LO}}%
}{E_{0}^{\text{free}}}+Ae^{-\delta E\cdot t}, \label{asymptotic1}%
\end{equation}%
\begin{equation}
\frac{\Delta E_{\text{NLO}}(t)}{E_{0}^{\text{free}}}\approx\frac
{E_{0,\text{NLO}}-E_{0,\text{LO}}}{E_{0}^{\text{free}}}+Be^{-\delta E\cdot
t/2}. \label{asymptotic2}%
\end{equation}%
\begin{equation}
\frac{E_{\text{LO}}(t)+\Delta E_{\text{NLO}}(t)}{E_{0}^{\text{free}}}%
\approx\frac{E_{0,\text{NLO}}}{E_{0}^{\text{free}}}+Ae^{-\delta E\cdot
t}+Be^{-\delta E\cdot t/2}. \label{asymptotic3}%
\end{equation}
The unknown coefficients $A$ and $B$, energy gap $\delta E$, and ground state
energies $E_{0,\text{LO}}$ and $E_{0,\text{NLO}}$ are determined by least
squares fitting. \ The $e^{-\delta E\cdot t}$ dependence in
Eq.~(\ref{asymptotic1}) comes from the contribution of the lowest excited
state with energy $\delta E$ above the ground state. The $e^{-\delta E\cdot
t/2}$ dependence in Eq.~(\ref{asymptotic2}) comes from the matrix element of
$M_{\text{NLO}}$ between the ground state and the lowest excited state. \ The
reduced chi-square for each fit is shown in Fig.~\ref{n8} and \ref{n12}, and
in each case they are close to 1.%
\ 

\begin{figure}
[ptb]
\begin{center}
\includegraphics[
height=3.224in,
width=3.6616in
]%
{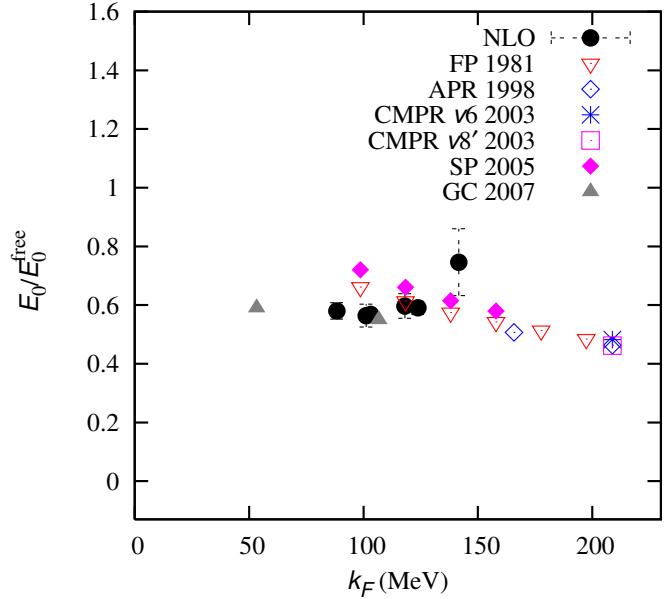}%
\caption{Results for $E_{0,\text{NLO}}/E_{\text{0}}^{\text{free}}$ versus
Fermi momentum $k_{F}$. \ For comparison we show the results for FP 1981
\cite{Friedman:1981qw}, APR 1998 \cite{Akmal:1998cf}, CMPR $v6$ and
$v8^{\prime}$ 2003 \cite{Carlson:2003wm}, SP 2005 \cite{Schwenk:2005ka}, and
GC 2007 \cite{Gezerlis:2007fs}.}%
\label{xsi_literature}%
\end{center}
\end{figure}

We calculate the Fermi momentum $k_{F}$ for each neutron spin from the
corresponding density. \ In our case $\rho_{\uparrow}=\rho_{\downarrow
}=N/(2L^{3})$ and so%
\begin{equation}
k_{F}=\frac{1}{L}\left(  3\pi^{2}N\right)  ^{1/3}.
\end{equation}
In Fig.~\ref{xsi_literature} we show the results for $E_{0,\text{NLO}%
}/E_{\text{0}}^{\text{free}}$ versus Fermi momentum $k_{F}$. \ The error bars
on $E_{0,\text{NLO}}/E_{\text{0}}^{\text{free}}$ represent uncertainties from
the asymptotic fits in Eq.~(\ref{asymptotic1})-(\ref{asymptotic3}). \ For
comparison we show other results from the literature: \ FP 1981
\cite{Friedman:1981qw}, APR 1998 \cite{Akmal:1998cf}, CMPR $v6$ and
$v8^{\prime}$ \cite{Carlson:2003wm}, SP 2005 \cite{Schwenk:2005ka}, and GC
2007 \cite{Gezerlis:2007fs}. \ We find good agreement near $k_{F}=120$ MeV but
there is disagreement whether the slope is positive or negative.%

\section{Analysis and discussion}

Neutron matter at $k_{F}\sim80$~MeV is close to the idealized unitarity limit,
where the $S$-wave scattering length is infinite and the range of the
interaction is negligible. \ At lower densities corrections due to the
scattering length become more important, and at higher densities corrections
due to the effective range and other effects become important. \ In the
unitarity limit the ground state has no dimensionful parameters other than
particle density and so the ground state energy of the system should obey a
simple relation $E_{0}=\xi E_{0}^{\text{free}}$ for some dimensionless
constant $\xi$. \ The universal nature of the unitarity limit endows it
relevance to several areas of physics, and in atomic physics the unitarity
limit has been studied extensively with ultracold $^{6}$Li and $^{40}$K atoms
using a magnetic-field Feshbach resonance
\cite{Tiesinga:1993PRA,Stwalley:1976PRL,Courteille:1998PRL,Inouye:1998Nat}.

Recent experiments for $\xi$ have measured the expansion of $^{6}$Li and
$^{40}$K in the unitarity limit released from a harmonic trap. \ The measured
values for $\xi$ are $0.51(4)$ \cite{Kinast:2005}, $0.46_{-05}^{+12}$
\cite{Stewart:2006}, and $0.32_{-13}^{+10}$ \cite{Bartenstein:2004}. \ The
discrepancy between these measurements and larger values for $\xi$ reported in
earlier experiments \cite{O'Hara:2002,Bourdel:2003,Gehm:2003} suggests that
further work may be needed.

There have been numerous analytic calculations of $\xi$
\cite{Engelbrecht:1997,Baker:1999dg,Heiselberg:1999,Perali:2004,Schafer:2005kg,Nishida:2006a,Nishida:2006b,Arnold:2007,Nikolic:2007,Veillette:2006}%
. \ The values for $\xi$ vary roughly from $0.2$ to $0.6$. \ Fixed-node
Green's function Monte Carlo calculations have found $\xi$ to be $0.44(1)$
\cite{Carlson:2003z} and $0.42(1)$ \cite{Astrakharchik:2004}. \ An estimate
based on Kohn-Sham theory for the two-fermion system in a harmonic trap yields
a value of $0.42$ \cite{Papenbrock:2005}. \ There have also been simulations
of two-component fermions on the lattice in the unitarity limit at non-zero
temperature. \ When data are extrapolated to zero temperature the results of
\cite{Bulgac:2005a} produce a value for $\xi$ similar to the fixed-node
results. \ The same is true for \cite{Burovski:2006a,Burovski:2006b}, though
with significant error bars, while calculations by Lee and Sch\"{a}fer
\cite{Lee:2005is,Lee:2005it} established a bound, $0.07\leq\xi\leq0.42$.

For finite $S$-wave scattering length $a_{0}$ the deviation away from
unitarity can be parameterized as%
\begin{equation}
\frac{E_{0}}{E_{\text{0}}^{\text{free}}}\approx\xi-\frac{\xi_{1}}{k_{F}a_{0}}.
\end{equation}
Both $\xi$ and $\xi_{1}$ have been computed using the lattice transfer matrix
projection method discussed above. \ The results are $\xi=0.25(3)$
\cite{Lee:2005fk} and $\xi_{1}=1.0(1)$ \cite{Lee:2006hr}. \ More recent
lattice calculations find similar values for $\xi$ and $\xi_{1}$
\cite{Lee:2007A,Abe:2007fe,Abe:2007ff}. \ There is general agreement in the
recent literature on the value of $\xi_{1}$
\cite{Chang:2004PRA,Astrakharchik:2004,Chen:2006A}. \ Further work will be
needed to resolve the remaining differences between the various calculations
for $\xi$. \ We use the values from \cite{Lee:2005fk} and \cite{Lee:2006hr} in
our analysis.

\begin{figure}
[ptb]
\begin{center}
\includegraphics[
height=2.4215in,
width=3.7766in
]%
{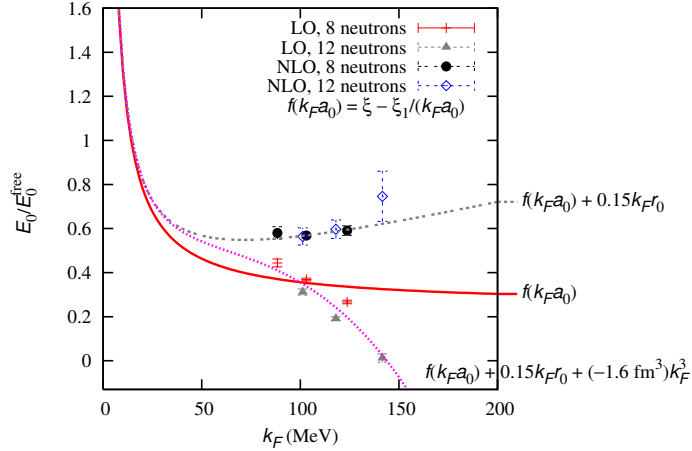}%
\caption{Plot of $E_{0,\text{LO}}/E_{\text{0}}^{\text{free}}$ and
$E_{0,\text{NLO}}/E_{\text{0}}^{\text{free}}$ versus $k_{F}$. \ For comparison
we plot $f(k_{F}a_{0}),$ $f(k_{F}a_{0})+0.15k_{F}r_{0}$, and $f(k_{F}%
a_{0})+0.15k_{F}r_{0}+(-1.6$ fm$^{3})k_{F}^{3}$.}%
\label{xsi_kf}%
\end{center}
\end{figure}

In addition to the corrections at finite scattering length we expect
corrections proportional to $k_{F}r_{0}$ due to the $S$-wave effective range
$r_{0}$. \ We also expect higher-order corrections away from the unitarity
limit arising from higher powers of $1/(k_{F}a_{0})$ and $k_{F}r_{0}$, as well
as other terms associated with the $S$-wave shape parameter and triplet
$P$-wave scattering volumes. \ In Fig.~\ref{xsi_kf} we show both
$E_{0,\text{LO}}/E_{\text{0}}^{\text{free}}$ and $E_{0,\text{NLO}}%
/E_{\text{0}}^{\text{free}}$ versus $k_{F}$. \ For comparison we plot%
\begin{equation}
f(k_{F}a_{0})=\xi-\frac{\xi_{1}}{k_{F}a_{0}}, \label{f}%
\end{equation}
with $\xi=0.25$, $\xi_{1}=1.0$, and neutron scattering length $a_{0}=-18.5$
fm. \ From Fig.~\ref{xsi_kf} we see that the NLO energy ratio $E_{0,\text{NLO}%
}/E_{\text{0}}^{\text{free}}$ is approximately described by
\begin{equation}
E_{0,\text{NLO}}/E_{\text{0}}^{\text{free}}\approx f(k_{F}a_{0})+0.15k_{F}%
r_{0}, \label{kf1}%
\end{equation}
where $r_{0}$ is the neutron effective range $2.7$ fm. \ The $k_{F}r_{0}$ term
in Eq.~(\ref{kf1}) can be interpreted as the correction due to the neutron
effective range. \ But as noted above there should also be corrections from
higher powers of $1/(k_{F}a_{0})$ and $k_{F}r_{0}\ $and from the $S$-wave
shape parameter and triplet $P$-wave scattering volumes. \ It is not obvious
why these higher-order effects are all numerically small at $k_{F}\approx
m_{\pi}$ as the NLO lattice results suggest.

In contrast we see deviations beyond the $1/(k_{F}a_{0})$ and $k_{F}r_{0}$
corrections in the LO\ lattice results. \ As shown in Fig.~\ref{xsi_kf} the
leading-order ratio $E_{0,\text{LO}}/E_{\text{0}}^{\text{free}}$ appears to
lie on the curve%
\begin{equation}
E_{0,\text{LO}}/E_{\text{0}}^{\text{free}}\approx f(k_{F}a_{0})+0.15k_{F}%
r_{0}+(-1.6\text{ fm}^{3})k_{F}^{3}. \label{kf3}%
\end{equation}%
We know from the $^{1}S_{0}$ phase shifts in Fig.~\ref{i1waves} that $S$-wave
scattering for the LO and NLO actions are nearly identical. \ This explains
the common coefficient of $0.15$ in front of $k_{F}r_{0}$ for both LO and
NLO\ results. \ Therefore, the difference between LO and NLO\ results must
come from interactions in higher partial waves.

For the LO action each of triplet $P$-wave interactions in Fig.~\ref{i1waves}
are attractive. \ The $(-1.6$ fm$^{3})k_{F}^{3}$ term in Eq.~(\ref{kf3}) for
the LO action is consistent with the type of correction we expect from the
negative triplet $P$-wave scattering volumes. \ On the other hand, the
$k_{F}^{3}$ correction from $P$-wave interactions in the NLO\ action seems to
be numerically very small. \ To understand this better we probe the relation
between low-energy $P$-wave interactions and the energy ratio $E_{0,\text{NLO}%
}/E_{\text{0}}^{\text{free}}$ by varying coefficients of the NLO operators.

\begin{figure}
[ptb]
\begin{center}
\includegraphics[
height=2.4249in,
width=3.1116in
]%
{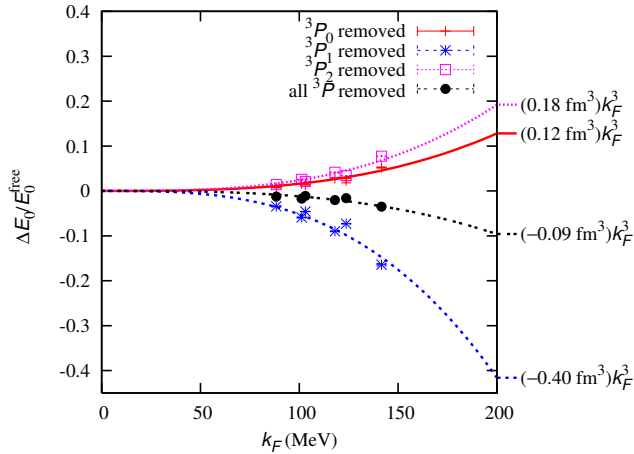}%
\caption{The change $\Delta E_{0,\text{NLO}}/E_{\text{0}}^{\text{free}}$ due
to removing $^{3}P_{0}$, $^{3}P_{1}$, $^{3}P_{2}$, or all triplet $P$-wave
interactions.}%
\label{xsi_kf_pwave_differences}%
\end{center}
\end{figure}

\begin{figure}
[ptb]
\begin{center}
\includegraphics[
height=2.4232in,
width=3.1946in
]%
{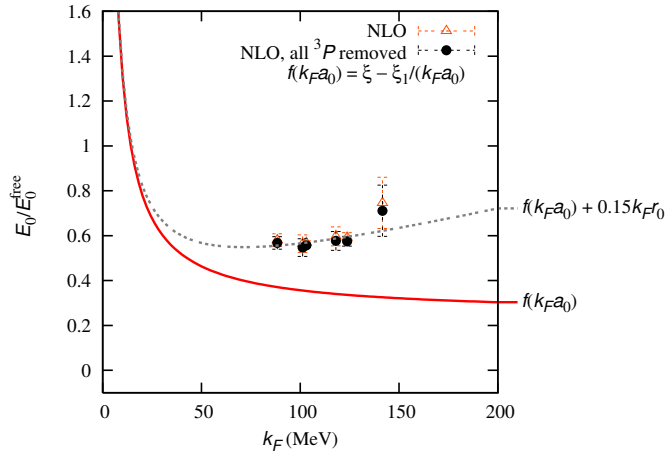}%
\caption{Plot of $E_{0,\text{NLO}}/E_{\text{0}}^{\text{free}}$ versus $k_{F}$
and the effect of removing all triplet $P$-wave contributions.}%
\label{xsi_kf_pwave}%
\end{center}
\end{figure}

The NLO coefficients in Table \ref{NLOcoefficients} were determined by fitting
the five data points labelled by arrows in Fig.~\ref{i1waves}. \ We consider
four variations of these NLO coefficients. \ For the first variation we set
the phase shift for the $^{3}P_{0}$ data point to zero while keeping other
data points the same. \ For the second variation we zero out the phase shift
of the $^{3}P_{1}$ data point while keeping others the same. \ For the third
variation we zero out only the $^{3}P_{2}$ phase shift, and for the fourth we
zero out all three triplet $P$-waves. \ The change $\Delta E_{0,\text{NLO}%
}/E_{\text{0}}^{\text{free}}$ due to each of these variations is plotted in
Fig.~\ref{xsi_kf_pwave_differences}. \ The results show significant
cancellation between the $P$-wave contributions. \ In fact the total
contribution from all $P$-waves is smaller than any individual contribution.
\
In Fig.~\ref{xsi_kf_pwave} we show $E_{0,\text{NLO}}/E_{\text{0}}%
^{\text{free}}$ and the effect of removing all triplet $P$-wave contributions.
\ Both data sets lie approximately on the curve%
\begin{equation}
E_{0,\text{NLO}}/E_{\text{0}}^{\text{free}}\approx f(k_{F}a_{0})+0.15k_{F}%
r_{0}.
\end{equation}
We see that the total effect of the triplet $P$-wave scattering volumes is
small due to cancellations between the $J=0,1,2$ contributions.%

\section{Summary}

We have discussed lattice simulations of the ground state of dilute neutron
matter using chiral effective field theory at next-to-leading order. \ In the
first paper the coefficients of the next-to-leading-order lattice action were
determined by matching nucleon-nucleon scattering data for momenta up to the
pion mass. \ In this second paper we used the same lattice action to simulate
the ground state of up to 12 neutrons in a periodic cube using Monte Carlo for
the density range from 2\% to 8\% of normal nuclear density. \ We found
agreement near $k_{F}=120$ MeV for the ground energy ratio $E_{0}/E_{\text{0}%
}^{\text{free}}$ with results in the literature. \ However there is
disagreement on whether the ratio is slightly increasing or slightly
decreasing with $k_{F}$.

We analyzed the energy ratio as an expansion about the unitarity limit with
corrections due to finite scattering length, effective range, and $P$-wave
interactions. \ We find significant cancellation between the various triplet
$P$-wave contributions to the ground state energy. \ We find a good fit to the
lattice data using
\begin{equation}
E_{0\text{,NLO}}/E_{\text{0}}^{\text{free}}\approx\xi-\frac{\xi_{1}}%
{k_{F}a_{0}}+0.15k_{F}r_{0},
\end{equation}
with $\xi=0.25$, $\xi_{1}=1.0$. \ The coefficient in front of $k_{F}r_{0}$
should be a universal constant and therefore measurable in other quantum
systems near the unitarity limit. \ In future studies we will consider larger
systems of dilute neutron matter and test model independence of results in the
manner discussed in \cite{FirstPaper}\ using several different lattice actions.

\section*{Acknowledgements}

Partial financial support from the Deutsche Forschungsgemeinschaft (SFB/TR
16), Helmholtz Association (contract number VH-NG-222 and VH-VI-231), and U.S.
Department of Energy (DE-FG02-03ER41260) are gratefully acknowledged. \ This
research is part of the EU Integrated Infrastructure Initiative in Hadron
Physics under contract number RII3-CT-2004-506078. \ The computational
resources for this project were provided by the John von Neumann Institute for
Computing at the Forschungszentrum J\"{u}lich.

\appendix

\section{Lattice action}

\subsection{Notation}

We assume exact isospin symmetry and neglect electromagnetic interactions.
\ $\vec{n}$ represents integer-valued lattice vectors on a three-dimensional
spatial lattice, and $\vec{p},$ $\vec{q},$ $\vec{k}$ represent integer-valued
momentum lattice vectors.$\ \ \hat{l}=\hat{1}$, $\hat{2}$, $\hat{3}$ are unit
lattice vectors in the spatial directions, $a$ is the spatial lattice spacing,
and $L$ is the length of the cubic spatial lattice in each direction. \ The
lattice time step is $a_{t}$, and $n_{t}$ labels the number of time steps.
\ We define $\alpha_{t}$ as the ratio between lattice spacings, $\alpha
_{t}=a_{t}/a$. \ Throughout we use dimensionless parameters and operators,
which correspond with physical values multiplied by the appropriate power of
$a$. \ Final results are presented in physical units with the corresponding
unit stated explicitly.

We use $a$ and $a^{\dagger}$ to denote annihilation and creation operators.
\ We make explicit all spin and isospin indices,%
\begin{align}
a_{0,0}  &  =a_{\uparrow,p},\text{ \ }a_{0,1}=a_{\uparrow,n},\\
a_{1,0}  &  =a_{\downarrow,p},\text{ \ }a_{1,1}=a_{\downarrow,n}.
\end{align}
The first subscript is for spin and the second subscript is for isospin. \ We
use $\tau_{I}$ with $I=1,2,3$ to represent Pauli matrices acting in isospin
space and $\sigma_{S}$ with $S=1,2,3$ to represent Pauli matrices acting in
spin space.

We use the eight vertices of a unit cube on the lattice to define spatial
derivatives. \ For each spatial direction $l=1,2,3$ and any lattice function
$f(\vec{n})$, let%
\begin{equation}
\Delta_{l}f(\vec{n})=\frac{1}{4}\sum_{\substack{\nu_{1},\nu_{2},\nu_{3}%
=0,1}}(-1)^{\nu_{l}+1}f(\vec{n}+\vec{\nu}),\qquad\vec{\nu}=\nu_{1}\hat{1}%
+\nu_{2}\hat{2}+\nu_{3}\hat{3}. \label{derivative}%
\end{equation}
We also define the double spatial derivative along direction $l$,%
\begin{equation}
\triangledown_{l}^{2}f(\vec{n})=f(\vec{n}+\hat{l})+f(\vec{n}-\hat{l}%
)-2f(\vec{n}). \label{d2l}%
\end{equation}

\subsection{Local densities and currents}

We define the local density,%
\begin{equation}
\rho^{a^{\dagger},a}(\vec{n})=\sum_{i,j=0,1}a_{i,j}^{\dagger}(\vec{n}%
)a_{i,j}(\vec{n}),
\end{equation}
which is invariant under Wigner's SU(4) symmetry \cite{Wigner:1937}.
\ Similarly we define the local spin density for $S=1,2,3,$%
\begin{equation}
\rho_{S}^{a^{\dagger},a}(\vec{n})=\sum_{i,j,i^{\prime}=0,1}a_{i,j}^{\dagger
}(\vec{n})\left[  \sigma_{S}\right]  _{ii^{\prime}}a_{i^{\prime},j}(\vec{n}),
\end{equation}
isospin density for $I=1,2,3,$%
\begin{equation}
\rho_{I}^{a^{\dagger},a}(\vec{n})=\sum_{i,j,j^{\prime}=0,1}a_{i,j}^{\dagger
}(\vec{n})\left[  \tau_{I}\right]  _{jj^{\prime}}a_{i,j^{\prime}}(\vec{n}),
\end{equation}
and spin-isospin density for $S,I=1,2,3,$%
\begin{equation}
\rho_{S,I}^{a^{\dagger},a}(\vec{n})=\sum_{i,j,i^{\prime},j^{\prime}%
=0,1}a_{i,j}^{\dagger}(\vec{n})\left[  \sigma_{S}\right]  _{ii^{\prime}%
}\left[  \tau_{I}\right]  _{jj^{\prime}}a_{i^{\prime},j^{\prime}}(\vec{n}).
\end{equation}

For each static density we also have an associated current density. \ Similar
to the definition of the lattice derivative $\Delta_{l}$ in (\ref{derivative}%
), we use the eight vertices of a unit cube,
\begin{equation}
\vec{\nu}=\nu_{1}\hat{1}+\nu_{2}\hat{2}+\nu_{3}\hat{3},
\end{equation}
for $\nu_{1},\nu_{2},\nu_{3}=0,1$. \ Let $\vec{\nu}(-l)$ for $l=1,2,3$ be the
result of reflecting the $l^{\text{th}}$-component of $\vec{\nu}$ about the
center of the cube,%
\begin{equation}
\vec{\nu}(-l)=\vec{\nu}+(1-2\nu_{l})\hat{l}.
\end{equation}
Omitting factors of $i$ and $1/m$, we can write the $l^{\text{th}}$-component
of the SU(4)-invariant current density as%
\begin{equation}
\Pi_{l}^{a^{\dagger},a}(\vec{n})=\frac{1}{4}\sum_{\substack{\nu_{1},\nu
_{2},\nu_{3}=0,1}}\sum_{i,j=0,1}(-1)^{\nu_{l}+1}a_{i,j}^{\dagger}(\vec{n}%
+\vec{\nu}(-l))a_{i,j}(\vec{n}+\vec{\nu}).
\end{equation}
Similarly the $l^{\text{th}}$-component of spin current density is%
\begin{equation}
\Pi_{l,S}^{a^{\dagger},a}(\vec{n})=\frac{1}{4}\sum_{\substack{\nu_{1},\nu
_{2},\nu_{3}=0,1}}\sum_{i,j,i^{\prime}=0,1}(-1)^{\nu_{l}+1}a_{i,j}^{\dagger
}(\vec{n}+\vec{\nu}(-l))\left[  \sigma_{S}\right]  _{ii^{\prime}}a_{i^{\prime
},j}(\vec{n}+\vec{\nu}),
\end{equation}
$l^{\text{th}}$-component of isospin current density is%
\begin{equation}
\Pi_{l,I}^{a^{\dagger},a}(\vec{n})=\frac{1}{4}\sum_{\substack{\nu_{1},\nu
_{2},\nu_{3}=0,1}}\sum_{i,j,j^{\prime}=0,1}(-1)^{\nu_{l}+1}a_{i,j}^{\dagger
}(\vec{n}+\vec{\nu}(-l))\left[  \tau_{I}\right]  _{jj^{\prime}}a_{i,j^{\prime
}}(\vec{n}+\vec{\nu}),
\end{equation}
and $l^{\text{th}}$-component of spin-isospin current density is%
\begin{equation}
\Pi_{l,S,I}^{a^{\dagger},a}(\vec{n})=\frac{1}{4}\sum_{\substack{\nu_{1}%
,\nu_{2},\nu_{3}=0,1}}\sum_{i,j,i^{\prime},j^{\prime}=0,1}(-1)^{\nu_{l}%
+1}a_{i,j}^{\dagger}(\vec{n}+\vec{\nu}(-l))\left[  \sigma_{S}\right]
_{ii^{\prime}}\left[  \tau_{I}\right]  _{jj^{\prime}}a_{i^{\prime},j^{\prime}%
}(\vec{n}+\vec{\nu}).
\end{equation}

\subsection{Instantaneous free pion action}

The lattice action for free pions with purely instantaneous propagation is%
\begin{equation}
S_{\pi\pi}(\pi_{I})=\alpha_{t}(\tfrac{m_{\pi}^{2}}{2}+3)\sum_{\vec{n},n_{t}%
,I}\pi_{I}(\vec{n},n_{t})\pi_{I}(\vec{n},n_{t})-\alpha_{t}\sum_{\vec{n}%
,n_{t},I,l}\pi_{I}(\vec{n},n_{t})\pi_{I}(\vec{n}+\hat{l},n_{t}),
\end{equation}
where $\pi_{I}$ is the pion field labelled with isospin index $I$. \ It is
convenient to define a rescaled pion field, $\pi_{I}^{\prime}$,%
\begin{equation}
\pi_{I}^{\prime}(\vec{n},n_{t})=\sqrt{q_{\pi}}\pi_{I}(\vec{n},n_{t}),
\end{equation}%
\begin{equation}
q_{\pi}=\alpha_{t}(m_{\pi}^{2}+6).
\end{equation}
Then%
\begin{equation}
S_{\pi\pi}(\pi_{I}^{\prime})=\frac{1}{2}\sum_{\vec{n},n_{t},I}\pi_{I}^{\prime
}(\vec{n},n_{t})\pi_{I}^{\prime}(\vec{n},n_{t})-\frac{\alpha_{t}}{q_{\pi}}%
\sum_{\vec{n},n_{t},I,l}\pi_{I}^{\prime}(\vec{n},n_{t})\pi_{I}^{\prime}%
(\vec{n}+\hat{l},n_{t}). \label{pionaction}%
\end{equation}

In momentum space the action is%
\begin{equation}
S_{\pi\pi}(\pi_{I}^{\prime})=\frac{1}{L^{3}}\sum_{I,\vec{k}}\pi_{I}^{\prime
}(-\vec{k},n_{t})\pi_{I}^{\prime}(\vec{k},n_{t})\left[  \frac{1}{2}%
-\frac{\alpha_{t}}{q_{\pi}}\sum_{l}\cos\left(  \tfrac{2\pi k_{l}}{L}\right)
\right]  .
\end{equation}
The instantaneous pion correlation function at spatial separation $\vec{n}$ is%
\begin{align}
\left\langle \pi_{I}^{\prime}(\vec{n},n_{t})\pi_{I}^{\prime}(\vec{0}%
,n_{t})\right\rangle  &  =\frac{\int D\pi_{I}^{\prime}\;\pi_{I}^{\prime}%
(\vec{n},n_{t})\pi_{I}^{\prime}(\vec{0},n_{t})\;\exp\left[  -S_{\pi\pi
}\right]  }{\int D\pi_{I}^{\prime}\;\exp\left[  -S_{\pi\pi}\right]  }\text{
\ (no sum on }I\text{)}\nonumber\\
&  =\frac{1}{L^{3}}\sum_{\vec{k}}e^{-i\frac{2\pi}{L}\vec{k}\cdot\vec{n}}%
D_{\pi}(\vec{k}),
\end{align}
where%
\begin{equation}
D_{\pi}(\vec{k})=\frac{1}{1-\tfrac{2\alpha_{t}}{q_{\pi}}\sum_{l}\cos\left(
\tfrac{2\pi k_{l}}{L}\right)  }.
\end{equation}
It is useful also to define the two-derivative pion correlator, $G_{S_{1}%
S_{2}}(\vec{n})$,%
\begin{align}
G_{S_{1}S_{2}}(\vec{n})  &  =\left\langle \Delta_{S_{1}}\pi_{I}^{\prime}%
(\vec{n},n_{t})\Delta_{S_{2}}\pi_{I}^{\prime}(\vec{0},n_{t})\right\rangle
\text{ \ (no sum on }I\text{)}\nonumber\\
&  =\frac{1}{16}\sum_{\nu_{1},\nu_{2},\nu_{3}=0,1}\sum_{\nu_{1}^{\prime}%
,\nu_{2}^{\prime},\nu_{3}^{\prime}=0,1}(-1)^{\nu_{S_{1}}}(-1)^{\nu_{S_{2}%
}^{\prime}}\left\langle \pi_{I}^{\prime}(\vec{n}+\vec{\nu}-\vec{\nu}^{\prime
},n_{t})\pi_{I}^{\prime}(\vec{0},n_{t})\right\rangle .
\end{align}

\subsection{Leading-order transfer matrix LO$_{2}$}

We use the $O(a^{4})$-improved free lattice Hamiltonian,%
\begin{align}
H_{\text{free}}  &  =\frac{49}{12m}\sum_{\vec{n}}\sum_{i,j=0,1}a_{i,j}%
^{\dagger}(\vec{n})a_{i,j}(\vec{n})\nonumber\\
&  -\frac{3}{4m}\sum_{\vec{n}}\sum_{i,j=0,1}\sum_{l=1,2,3}\left[
a_{i,j}^{\dagger}(\vec{n})a_{i,j}(\vec{n}+\hat{l})+a_{i,j}^{\dagger}(\vec
{n})a_{i,j}(\vec{n}-\hat{l})\right] \nonumber\\
&  +\frac{3}{40m}\sum_{\vec{n}}\sum_{i,j=0,1}\sum_{l=1,2,3}\left[
a_{i,j}^{\dagger}(\vec{n})a_{i,j}(\vec{n}+2\hat{l})+a_{i,j}^{\dagger}(\vec
{n})a_{i,j}(\vec{n}-2\hat{l})\right] \nonumber\\
&  -\frac{1}{180m}\sum_{\vec{n}}\sum_{i,j=0,1}\sum_{l=1,2,3}\left[
a_{i,j}^{\dagger}(\vec{n})a_{i,j}(\vec{n}+3\hat{l})+a_{i,j}^{\dagger}(\vec
{n})a_{i,j}(\vec{n}-3\hat{l})\right]  .
\end{align}
The leading-order transfer matrix designated $M_{\text{LO}_{2}}$ in
\cite{Borasoy:2006qn} is%
\begin{align}
M_{\text{LO}_{2}}  &  =\colon\exp\left\{  -H_{\text{free}}\alpha_{t}%
-\frac{\alpha_{t}}{2L^{3}}\sum_{\vec{q}}f(q^{2})\left[  C\rho^{a^{\dag}%
,a}(\vec{q})\rho^{a^{\dag},a}(-\vec{q})+C_{I^{2}}\sum_{I}\rho_{I}^{a^{\dag}%
,a}(\vec{q})\rho_{I}^{a^{\dag},a}(-\vec{q})\right]  \right. \nonumber\\
&  +\left.  \frac{g_{A}^{2}\alpha_{t}^{2}}{8f_{\pi}^{2}q_{\pi}}\sum
_{\substack{S_{1},S_{2},I}}\sum_{\vec{n}_{1},\vec{n}_{2}}G_{S_{1}S_{2}}%
(\vec{n}_{1}-\vec{n}_{2})\rho_{S_{1},I}^{a^{\dag},a}(\vec{n}_{1})\rho
_{S_{2},I}^{a^{\dag},a}(\vec{n}_{2})\right\}  \colon.
\end{align}
where the momentum-dependent coefficient function $f(q^{2})$ is defined as
\begin{equation}
f(q^{2})=f_{0}^{-1}\exp\left[  -b%
{\displaystyle\sum\limits_{l}}
\left(  1-\cos q_{l}\right)  \right]  ,
\end{equation}
and the normalization factor $f_{0}$ is determined by the condition%
\begin{equation}
f_{0}=\frac{1}{L^{3}}\sum_{\vec{q}}\exp\left[  -b%
{\displaystyle\sum\limits_{l}}
\left(  1-\cos q_{l}\right)  \right]  .
\end{equation}
The value $b=0.6$ gives approximately the correct average effective range for
the two $S$-wave channels when $C$ and $C_{I^{2}}$ are properly tuned. \ $C$
is the coefficient of the Wigner SU(4)-invariant contact interaction and
$C_{I^{2}}$ is the coefficient of the isospin-dependent contact interaction.
$\ $For $C$ and $C_{I^{2}}$ we use the values%
\begin{equation}
C=\left(  3C^{I=1}+C^{I=0}\right)  /4, \label{C_coeff}%
\end{equation}%
\begin{equation}
C_{I^{2}}=\left(  C^{I=1}-C^{I=0}\right)  /4, \label{C_I2_coeff}%
\end{equation}
with $C^{I=1}=-3.414\times10^{-5}$ MeV$^{-2}$ and $C^{I=0}=-4.780\times
10^{-5}$ MeV$^{-2}$.

\bibliographystyle{apsrev}
\bibliography{NuclearMatter}

\end{document}